# Object Rigidity: Competition and cooperation between motion-energy and feature-tracking mechanisms and shape-based priors

Akihito Maruya & Qasim Zaidi*

Graduate Center for Vision Research, State University of New York, 33 West 42nd St, New York, NY 10036.

*Corresponding author: qz@sunyopt.edu

**AUTHOR CONTRIBUTIONS**

AM and QZ designed the study. AM programmed and ran the experiments. AM & QZ analyzed and modeled the results. AM & QZ wrote the paper.

The authors declare no competing interests.

## Abstract

Why do moving objects appear rigid when projected retinal images are deformed non-rigidly? We used rotating rigid objects that can appear rigid or non-rigid to test whether shape features contribute to rigidity perception. When two circular rings were rigidly linked at an angle and jointly rotated at moderate speeds, observers reported that the rings wobbled and were not linked rigidly but rigid rotation was reported at slow speeds. When gaps, paint or vertices were added, the rings appeared rigidly rotating even at moderate speeds. At high speeds, all configurations appeared non-rigid. Salient features thus contribute to rigidity at slow and moderate speeds, but not at high speeds. Simulated responses of arrays of motion-energy cells showed that motion flow vectors are predominantly orthogonal to the contours of the rings, not parallel to the rotation direction. A convolutional neural network trained to distinguish flow patterns for wobbling versus rotation, gave a high probability of wobbling for the motion-energy flows. However, the CNN gave high probabilities of rotation for motion flows generated by tracking features with arrays of MT pattern-motion cells and corner detectors. In addition, circular rings can appear to spin and roll despite the absence of any sensory evidence, and this illusion is prevented by vertices, gaps, and painted segments, showing the effects of rotational symmetry and shape. Combining CNN outputs that give greater weight to motion energy at fast speeds and to feature tracking at slow, with the shape-based priors for wobbling and rolling, explained rigid and nonrigid percepts across shapes and speeds ($R^2$=0.95). The results demonstrate how cooperation and competition between different neuronal classes lead to specific states of visual perception and to transitions between the states.

## Introduction

Visual neuroscience has been quite successful at identifying specialized neurons as the functional units of vision. Neuronal properties are important building blocks, but there's a big gap between understanding which stimuli drive a neuron and how cooperation and competition between different types of neurons generates visual perception. We try to bridge the gap for the perception of object rigidity and non-rigidity. Both of those states can be stable, so we need to develop ways of understanding how the visual system changes from one state to another. Shifts between different steady states are the general case for biological vision but have barely been investigated.

In the video of Figure 1A, most observers see the top ring as rolling or wobbling over the bottom ring, seemingly defying physical plausibility. In the video of Figure 1B, the two rings seem to be one rigid object rotating in a physically plausible way. Since both videos are of the same object rotating at different speeds, clearly an explanation is needed.

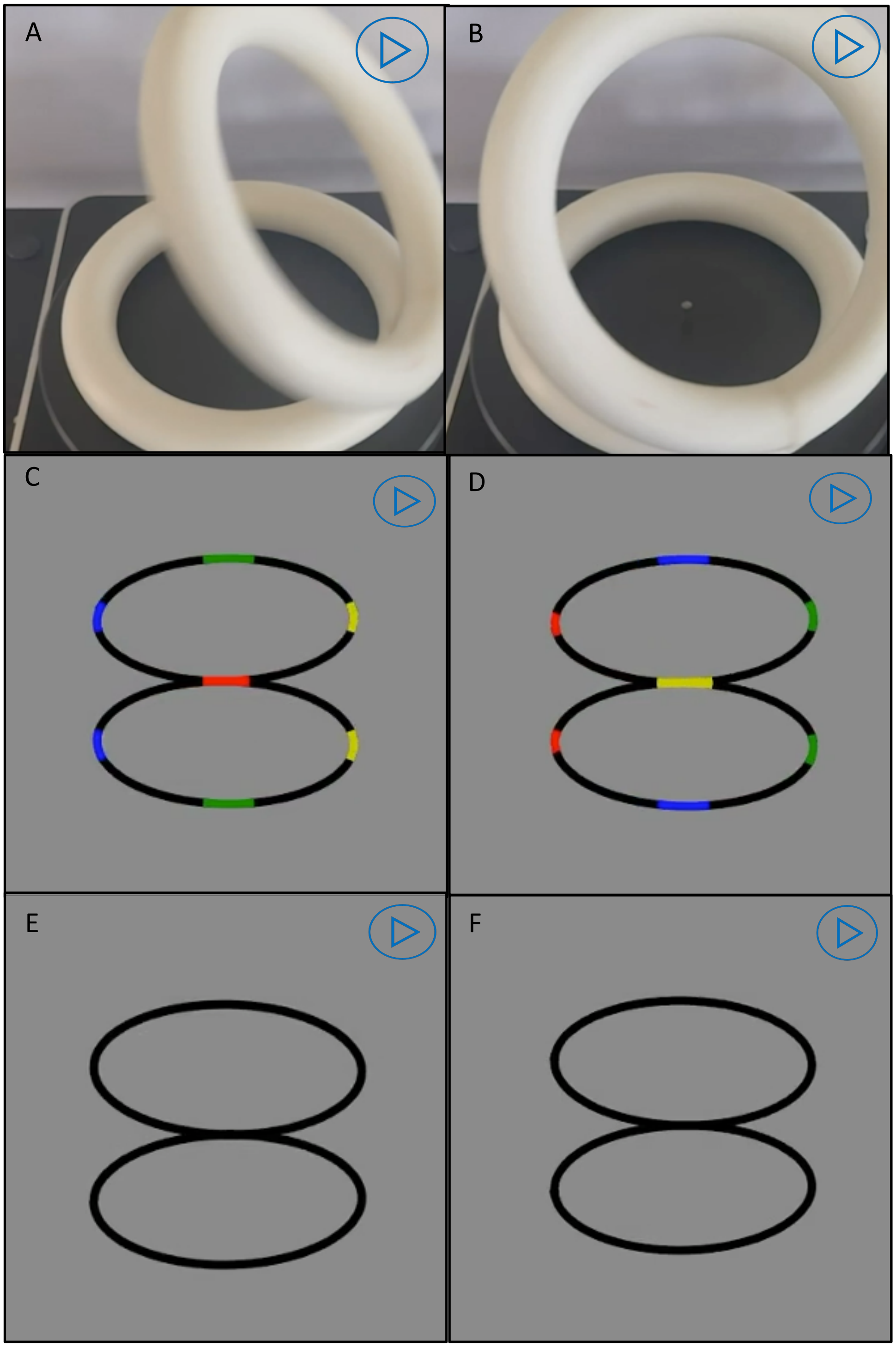

**Figure 1. Rotating ring illusion**. (A) A Styrofoam ring is seen at an angle to another ring as wobbling or rolling over the bottom ring despite physical implausibility. (B) At a slower speed, the rings are seen to rotate together revealing that they are glued together, and that the non-rigid rolling was an illusion. (C) Two rings rotate together with a fixed connection at the red segment. (D) Two rings wobble against each other shown by the connection shifting to other colors. (E) & (F) are the same as C & D except that the colored segments have been painted black. Both pairs generate identical sequence of retinal images so that wobbling and rotation are indistinguishable.

Humans can sometimes perceive the true shape of a moving object from impoverished information. For example, shadows show the perspective projection of just the object’s silhouette, yet 2D shadows can convey the 3D shape for some simple rotating objects, (Wallach and O'Connell,1953; Albertazzi, 2004). However, for irregular or unfamiliar objects when the light source is oblique to the surface on which the shadow is cast, shadows often get elongated and distorted in a way that the casting object is not recognizable. Similarly, from the shadow of an object passing rapidly, it is often difficult to discern whether the shadow is distorting or the object. Images on the retinae are also formed by perspective projection, and they too distort if the observer or the object is in motion, yet observers often correctly see the imaged object as rigid or nonrigid. Examples of rigidity often contain salient features, whereas rigid shapes without salient features are sometimes seen as non-rigid (Weiss and Adelson, 2000). We examine how and why salient features help in veridical perception of rigidity when objects are in motion. For that purpose, we use variations of the rigid object in Figures 1A & B that can appear rigid or non-rigid depending on speed and salience of features. This object is simple by standards of natural objects, but complex compared to stimuli generally used in studies of motion mechanisms. The interaction of motion with shape features has received extensive attention (McDermott et al., 2001; Papathomas, 2002; McDermott and Adelson, 2004; Berzhanskaya et al., 2007; Erlikhman et al., 2018). By exploring the competition and cooperation between motion energy mechanisms, feature tracking mechanisms, and shape-based priors, we present a mechanistic approach to perception of object rigidity.

To introduce controlled variation in the ring-pair, we switch from physical objects to computer graphics. Consider the videos of the two ring-pairs in Figures 1C & D. When attention is directed to the connecting joint, the configuration in Figure 1C is seen as rotating rigidly while the rings in Figure 1D wobble and slide non-rigidly against each other. Rigidity and non-rigidity are both easy to see because the painted segments at the junction of the two rings either stay together or slide past each other. The two videos in Figure 1E and F are of the same object as the two above, except that the painted segments have been turned to black. Now it is difficult to see any difference in the motion of the two rings because they form identical images, so whether they are seen as rigidly rotating or non-rigidly wobbling depends on how the images are processed by the visual system. The wobbling illusion of the rigidly connected rings has been used for many purposes. The senior author first saw it over 30 years ago at a tire shop in Texas, where it looked like a stationary horizontal tire was mounted on a high pole and a tire was rolling over it at an acute angle seemingly defying physical laws. The first author remembered that the wobbling rings illusion was used in the Superman movie to confine criminals during a trial. There are many videos on the internet showing how to make physical versions of the illusion, but despite the popularity of the illusion, we could not find an explanation of why people see non-rigidity in this rigid object.

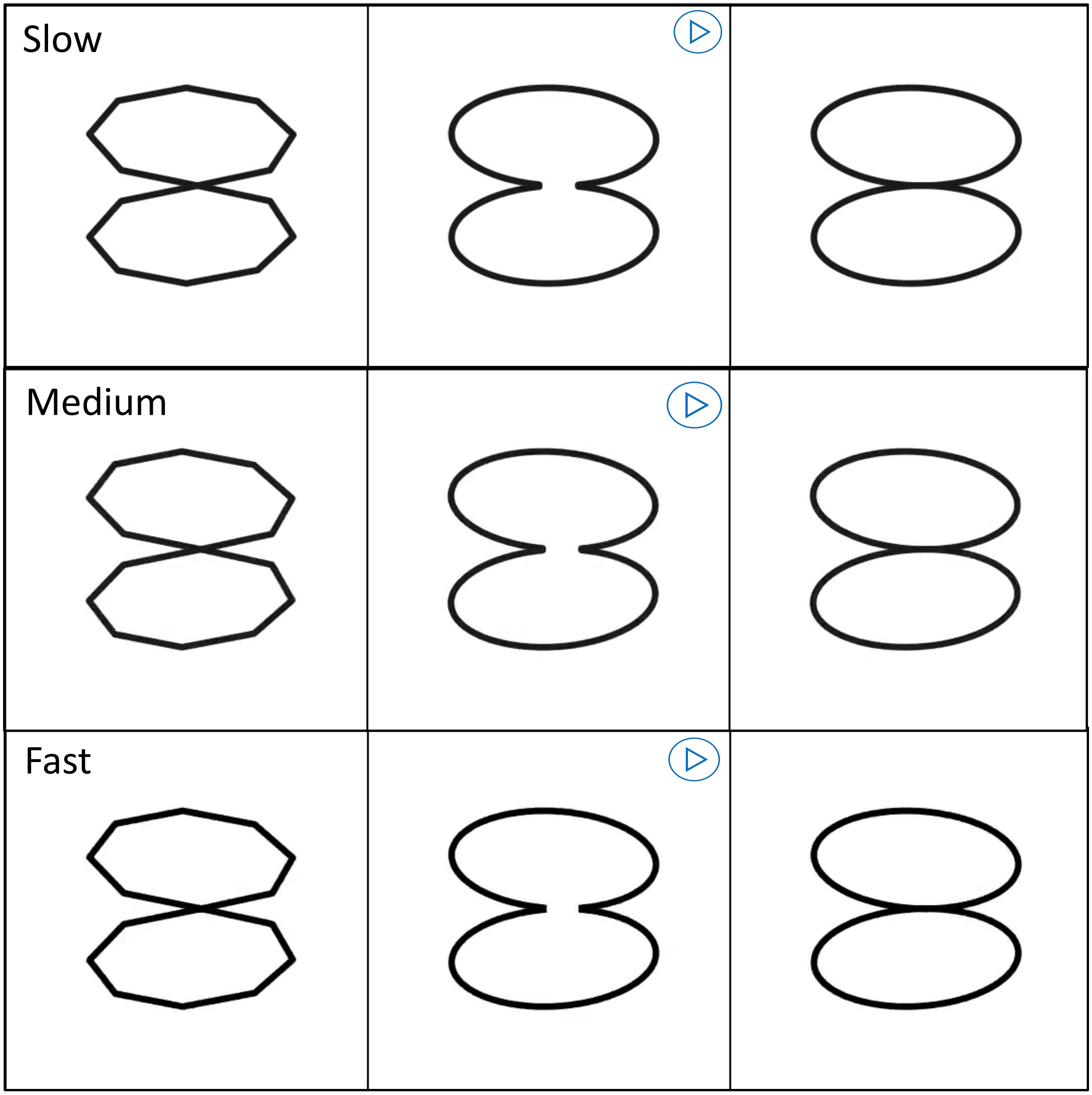

**Figure 2. Effect of shape on the ring illusion.** Pairs of rings with different shapes rotated around a vertical axis. When the speed is slow (1 deg/sec), all three shapes are seen as rotating. At medium speed (10 deg/sec), the circular rings seem to be wobbling against each other, but the other two shapes seem rigid. At fast speeds (30 deg/sec), nonrigid percepts dominate irrespective of shape features.

By varying the speed of rotation, we discovered that at slow speeds both the painted and non-painted rings appear rigidly connected and at high speeds both appear nonrigid. This was reminiscent of the differences in velocity requirements of motion-energy and feature-tracking mechanisms (Lu and Sperling, 1995; Zaidi and DeBonet, 2000), where motion-energy mechanisms function above a threshold velocity and feature-tracking only functions at slow speeds. Consequently, we processed the stimuli through arrays of motion-energy and feature-tracking units, and then trained a convolutional neural network to classify their outputs, demonstrating that the velocity-based relative importance of motion-energy versus feature tracking could explain the change in percepts with speed. To critically test this hypothesis, we manipulated the shape of the rings to create salient features that could be tracked more easily and used physical measures such as rotational symmetry to estimate prior expectations for wobbling and rolling of different shapes. These manipulations are demonstrated in Figure 2 where at medium speeds the rings with vertices and gaps appear rigidly rotating while the circular rings appear to be wobbling, whereas all three appear rigid at slow speeds, and non-rigidly connected at fast speeds (We established that the multiple images seen at fast speeds were generated by the visual system and were not in the display by taking photographs using a Canon T7 with a 1/6400 second shutter-speed). Taken together our investigations revealed previously unrecognized roles for feature-tracking and priors in maintaining veridical percepts of object rigidity.

## Perceived non-rigidity of a rigidly connected object

Shape from motion models generally invoke rigidity priors (Ullman 1979; Andersen and Bradley,1998). Besides the large class of rigid objects, there is also a large class of articulated objects in the world, including most animals whose limbs and trunks change shape for performing actions. If priors reflect statistics of the real world, it is quite likely that there is also a prior for objects consisting of connected parts to appear non-rigidly connected while parts appear rigid or at most elastic (Jain and Zaidi, 2011). This prior could support percepts of non-rigidity even when connected objects move rigidly, for example the rigid rotation of the ring pair. To quantify perception of the non-rigid illusion, we measured the proportion of times different shapes of ring pairs look rigidly or non-rigidly connected at different rotation speeds.

*Methods*

Using Python, we created two circular rings with a rigid connection at an angle. The rigid object rotates around a vertical axis oblique to both rings. The videos in Figure 3 show the stimuli that were used in this study to test the role of features in object rigidity. There were nine different shapes. The original circular ring pair is called “Circ ring”. A gap in the junction is called “Circ w gap”. The junction painted red is called “Circ w paint”. Two octagons were rigidly attached together at an edge “Oct on edge” or at a vertex “Oct on vertex”. Two squares were attached in the same manner “Sqr on vertex” and “Sqr on edge”. The junction between two ellipses (ratio of the longest to shortest axis was 4:3) was parallel to either the long or the short axis leading to “Wide ellipse” and “Long ellipse” respectively. A tenth configuration of the circular rings physically wobbling “Circ wobble”, was the only non-rigid configuration.

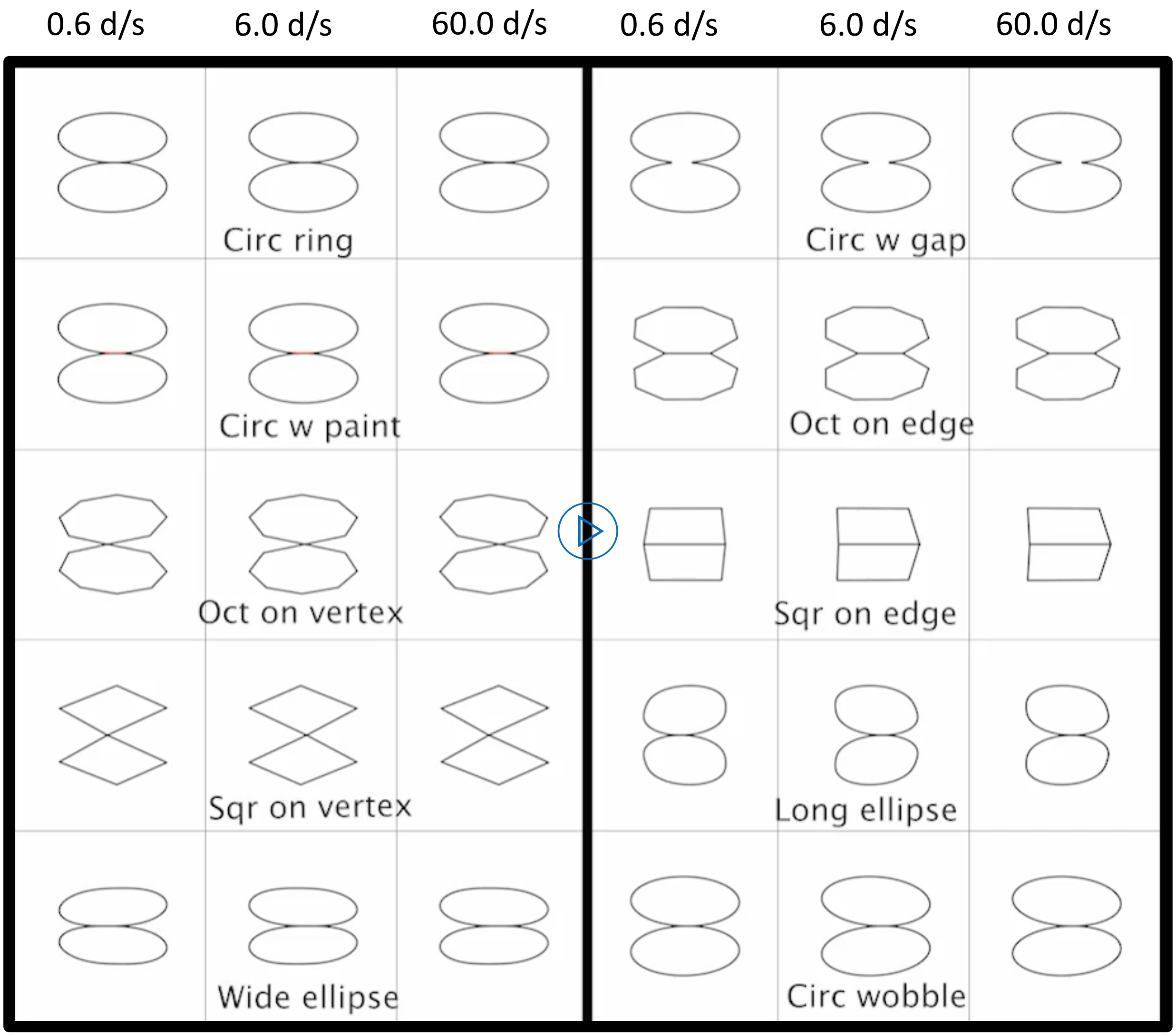

**Figure 3. Shapes showing the effects of features on object rigidity**. Rows give the names of the shapes. Columns give the three speeds.

The rotating ring pairs were rendered as if captured by a camera at a distance of 1.0 m and at either the same height as the junction ($0^o$ elevation) or at $15^o$ elevation (Equations in the **Appendix: Stimulus generation and projection**). We varied the rotation speed, but linear speed is more relevant for motion selective cells, so the speed of the joint when passing fronto-parallel was 0.6 dps, 6 dps, or 60 dps (degrees per second). The diameters of the rings were 3 dva or 6 dva (degrees of visual angle), so the corresponding angular speeds of rotation were 0.03, 0.3 and 3 cps (cycles per second) for the 3 dva rings and 0.015, 0.15, and 1.5 cps for the 6 dva rings. 3D cues other than motion and contour were eliminated by making the width of the line uniform and removing shading cues. In the videos of the stimuli (Figure 3), each row indicates a different shape of the stimulus, and the column represents the speed of the stimulus from 0.6 deg/sec (left) to 60 deg/sec (right).

The videos were displayed on a VIEWPixx/3D at 120 Hz. Matlab and PsychToolbox were used to display the stimulus and run the experiment. The data were analyzed using Python and Matlab. The initial rotational phase defined by the junction location was randomized for each trial as was the rotational direction (clockwise or counterclockwise looking down at the rings). An observer's viewing position was fixed by using a chin-rest so that the video was viewed at the same elevation as the camera position. The observer was asked to look at the junction between the rings and to report by pressing buttons if the rings were rigidly connected or not. The set of 120 conditions (10 shapes x 2 sizes x 3 speeds x 2 rotation directions) was repeated 20 times (10 times at each viewing elevation). Measurements were made by ten observers with normal or corrected to normal vision. Observers gave written informed consent. All experiments were conducted in compliance with a protocol approved by the institutional review board at SUNY College of Optometry, in compliance with the Declaration of Helsinki.

*Results*

Figures 4A & B show the average proportion of non-rigid percepts for each observer for each shape at the three speeds (0.6, 6.0 and 60.0 dps) for the 3 dva and 6 dva diameter sizes. Different colors indicate different observers, and the symbols are displaced horizontally to avoid some of them becoming hidden, the dark cross represents the mean. The results for the two sizes are similar, except that there is a slightly greater tendency to see rigidity for the larger size. The combined results for the two sizes, averaged over the 10 observers are shown as histograms in Figure 4 C. For the circular rings, there is a clear progression towards non-rigid percepts as the speed increases: At 0.6 dps a rigid rotation is perceived on average around 25% of the time. As the speed of rotation is increased, the average proportions of non-rigid percepts increase to around 60% at 6.0 dps and around 90% at 60.0 dps. These results provide empirical corroboration for the illusory non-rigidity of the rigidly rotating rings. Introducing a gap or painted segment in the circular rings increases the percept of rigidity especially at the medium speed. Turning the circular shapes into octagons with vertices further increases rigidity percepts and making the rings squares almost completely abolishes non-rigid percepts. If the circular rings are stretched into long ellipses that too reduces non-rigid percepts, but if they are stretched into wide ellipses, it has little effect, possibly because perspective shortening makes the projections of the ellipses close to circular. The results are averaged for the 10 observers in the histograms in Figure 4C, which will be used as a comparison with the model simulations. For all configurations other than the squares, non-rigid percepts increase as a function of increasing speed. The effect of salient features is thus greater at the slower speeds. The effect of speed provides clues for modeling the illusion based on established mechanisms for motion-energy and feature-tracking. Figure 4D shows the similarity between the results for the rotating and wobbling circular rings as the dots are close to the unit diagonal and $R^2 = 0.97$, which is not surprising given that their images are identical. Figure 4E shows that there is a slight tendency to see more non-rigidity at the 15° viewing elevation than the 0° elevation, but the $R^2 = 0.90$ meant that we could combine the data from the two viewpoints in the figures for simplicity.

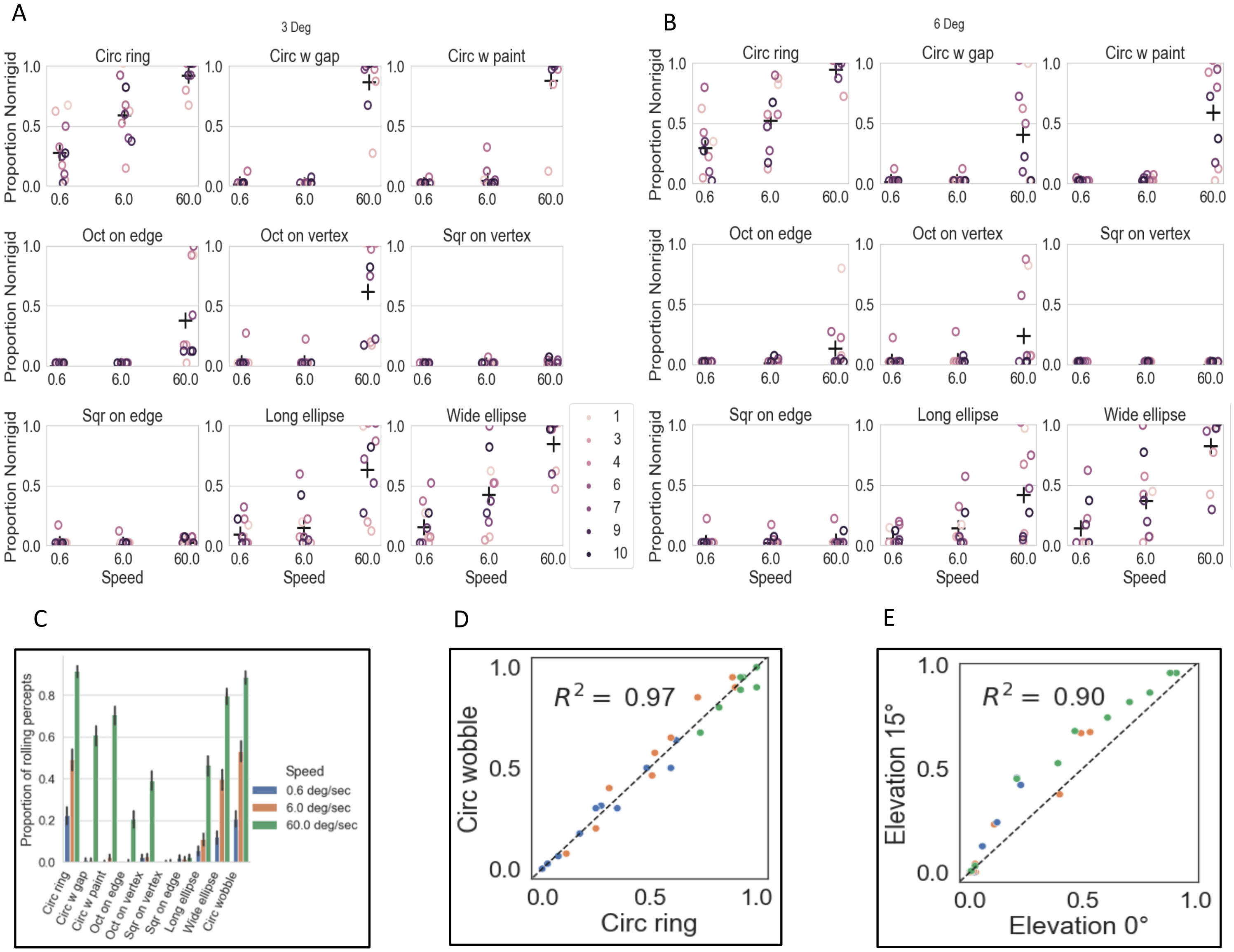

**Figure 4. Non-rigid percepts**. Average proportions of reports of non-rigidity for each of 10 observers for each shape at three speeds (0.6, 6.0 & 60.0 dps) for diameters of 3 dva (A) and 6 dva (B). Different colored circles indicate different observers and the average of all the observers is shown by the black cross. (C) Histograms of non-rigid percepts averaged over the 10 observers. (D) Average proportion of non-rigid percepts for the rotating and wobbling circular rings for 10 observers and 3 speeds. Similarity is shown by closeness to the unit diagonal and $R^2 = 0.97$. (E) Average proportion of non-rigid percepts for 0° elevation versus 15° elevation. Proportions are similar ($R^2 = 0.90$) , but slightly higher for the 15° elevation.

## Motion energy computations

The first question that arises is why observers see non-rigid motion when the physical motion is rigid. To get some insight into the answer, we simulated responses of direction-selective motion cells in primate primary visual cortex (Hubel 1959; Hubel and Wiesel, 1962; Movshon and Tolhurst, 1978a; Movshon and Tolhurst, 1978b) with a spatiotemporal energy model (Watson and Ahumada,1983; Adelson and Bergen, 1985; Watson and Ahumada, 1985; Van Santen and Sperling,1985; Nishimoto and Gallant, 2011; Nishimoto et al., 2011, Bradley and Goyal, 2008, Rust et al., 2006). A schematic diagram of a motion energy filter is shown in Figure 5A, and the equations are presented in the **Appendix: Motion Energy**. At the linear filtering stage, two spatially and temporally oriented filters in the quadrature phase were convolved with the sequence of images. Pairs of quadrature filters were squared and added to give phase invariant motion energy. Responses of V1 direction-selective cells are transmitted to extrastriate area MT, where cells include component and pattern cells (Movshon et al., 1985; Movshon and Newsome, 1996; Rust et al., 2006). Component cells have larger receptive fields than V1 direction-selective cells, but their motion responses are similar.

Each motion energy unit was composed of 2,000 direction-selective cells with five temporal frequencies (0, 2, 4, 6, and 8 Hz) times five spatial frequencies (0.2°, 0.3°, 0.375°, 0.75°, and 1.5°) times 16 directions (from 0° to 337.5° every 22.5°) times 5 sizes (Figure 5B). An array of 310248 motion energy units uniformly covering the whole stimulus were applied to the change at each pair of frames (Figure 5C). Since at every location many direction-selective cells respond at every instant, the responses have to be collapsed into a representation to visualize the velocity response dynamically. We use a color-coded vector whose direction and length at a location and instant respectively depict the preferred velocity of the unit that has the maximum response, akin to a winner-take-all rule (note that the length of the vector is not the magnitude of the response but the preferred velocity of the most responsive unit).

We begin by analyzing the circular ring pair since that shows the most change from rigid to non-rigid. Figure 5D indicates the motion energy field when the circular ring pair is rigidly rotating and Figure 5E shows when the two rings are physically wobbling. There are 200 x 200 x 199 (height × width × time frame) = 7960000 2-D vectors in each video. In both cases, for most locations and times, the preferred velocities are perpendicular to the contours of the rings. The response velocities in the rotating and wobbling rings look identical. To confirm this, Figure 5F subtracts 5E from 5D and shows that the difference is negligible. This vector field could thus contribute to the perception of wobbling or rotating or even be bistable. Since the vector directions are mostly not in the rotation direction, it would seem to support a percept of wobbling, but to provide a more objective answer, we trained a convolutional neural network (CNN) to discern between rotation and wobbling and fed it the motion energy vector field to perform a classification.

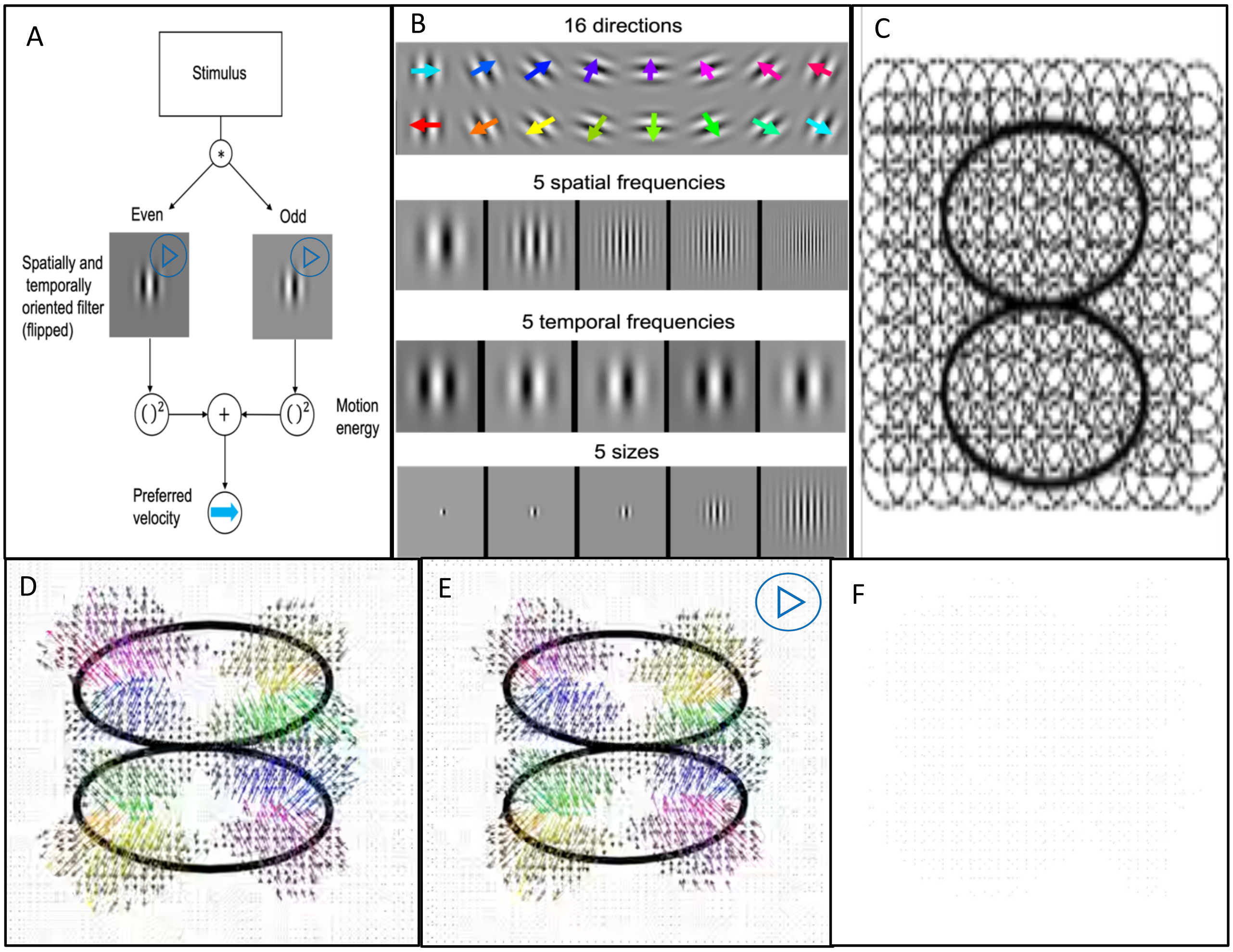

**Figure 5. Motion-energy mechanism**: (A) Schematic diagram of a motion energy unit: Moving stimulus is convolved with to filters that are odd and even symmetric spatially and temporally oriented filters, then the outputs are squared and summed to create a phase-independent motion energy response. (B) Motion energy units used in the model. At each spatial location there were 16 preferred directions, 5 spatial frequencies, 5 temporal frequencies, and 5 sizes. (C) An array of 310248 motion energy units uniformly covering the whole stimulus were applied to the change at each pair of frames. At each location, the preferred velocity of the highest responding motion energy unit from the array was selected as the population response. Motion vectors from physically rotating (D) and wobbling (E) ring pairs are predominantly orthogonal to contours instead of in the rotation direction. (F) The difference between the two vector fields is negligible. Since the flows for physically rotating and wobbling circular rings are almost identical, other factors must govern the perceptual shift from wobbling to rotation at slower speeds.

**Motion-pattern recognition Convolutional Neural Network**

For training the CNN, we generated random moving dot stimuli from a 3D space, and these dots either rotated around a vertical axis or wobbled at a random speed (0.1-9 deg/sec). The magnitude of wobbling was selected from -50º to 50º and the top and the bottom parts wobbled against each other with a similar magnitude as the wobbling ring stimulus that was presented to the observers. These 3D motions were projected to the 2D screen with camera elevations ranging from -45º to 45º. At each successive frame, the optimal velocity at each point was computed and the direction perturbed by Gaussian noise with sigma=1° to simulate noisy sensory evidence. Two examples of the 9000 vector fields (4500 rotational and 4500 wobbling) are shown in Figure 6A. As the examples show, it is easy for humans to discern the type of motion. The 9000 vector fields were randomly divided into 6300 training fields and 2700 validation fields.

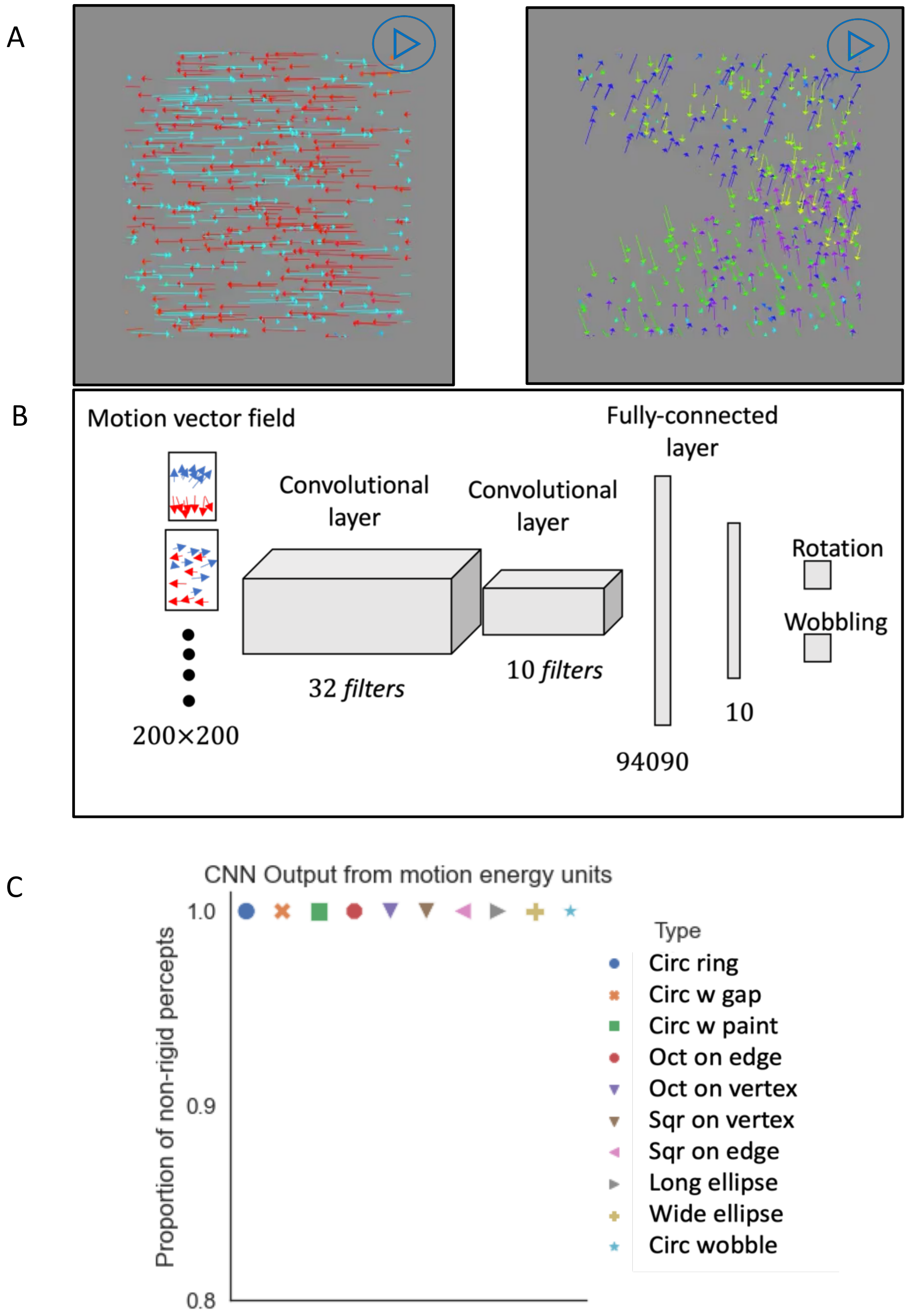

**Figure 6. Convolutional Neural Network for classifying patterns of motion vectors as rotating or wobbling.** (A) Two examples of the 9000 vector fields from random dot moving stimuli that were used to train and validate the CNN, (Left) the rotating vector field and (Right) the wobbling vector field. The 9000 vector fields were randomly divided into 6300 training and 2700 validation fields. (B) The network consists of two convolutional layers followed by two fully connected layers. The output layer gives a confidence level between rotation and wobbling on a 0.0 -1.0 scale. (C) Proportion of non-rigid percepts for CNN output from motion energy units for each shape. Symbol shape and color indicate ring-pair shape. For all ring shapes, the proportion of non-rigid classifications was 0.996.

The neural network was created and trained with TensorFlow (Abadi et al., 2016). For each pair of frames, the motion field is fed to the CNN as vectors (Figure 6B). The first layer of the CNN contains 32 filters each of which contain two channels for the horizontal and the vertical components of vectors. These filters are cross-correlated with the horizontal and vertical components of the random-dot flow fields. Then, each of the outputs is rectified and max pooled. The second layer contains 10 filters. Their output is flattened into arrays and followed by two fully connected layers that act like Perceptrons (Gallant, 1990). For each pair of frames, the last layer of the CNN provides a relative confidence level for wobbling versus rotation calculated by the softmax activation function (Sharma et al., 2017). Based on the higher confidence level for the set of frames in a trial, the network classifies the motion as rotation or wobbling (**Appendix: Convolutional Neural Network (CNN).)** After training for 5 epochs, the CNN reached 99.78 % accuracy for the training data set and 99.85 % for the validation test data set. We use the CNN purely as a pattern recognizer without any claims to biological validity, but its output does resemble position-independent neural responses to the pattern of the velocity field in cortical areas MST or STS (Tanaka et al., 1986; Sakata et al., 1986; Duffy and Wurtz, 1991; Zhang et al., 1993; Pitzalis et al., 2010).

We calculated motion energy vector fields for rotations of all 9 shapes and classified them with the CNN. At each time frame, the network reports a confidence level from 0 to 1 between wobbling and rotation. The proportion of non-rigid percepts of the CNN is derived by the average of the confidence level for wobbling across time frames. Classification of the motion-energy vectors lead to 99.6% percepts of wobbling for all the shapes at all the speeds (Figure 6C). This raises the question as to what makes observers perceive rotation for slow speeds, and different proportions of wobble for different shapes at medium speeds. To understand these differences from the motion-energy based predictions, we will examine feature tracking mechanisms, motion illusions that are unsupported by sensory signals, and prior assumptions based on shape and geometry.

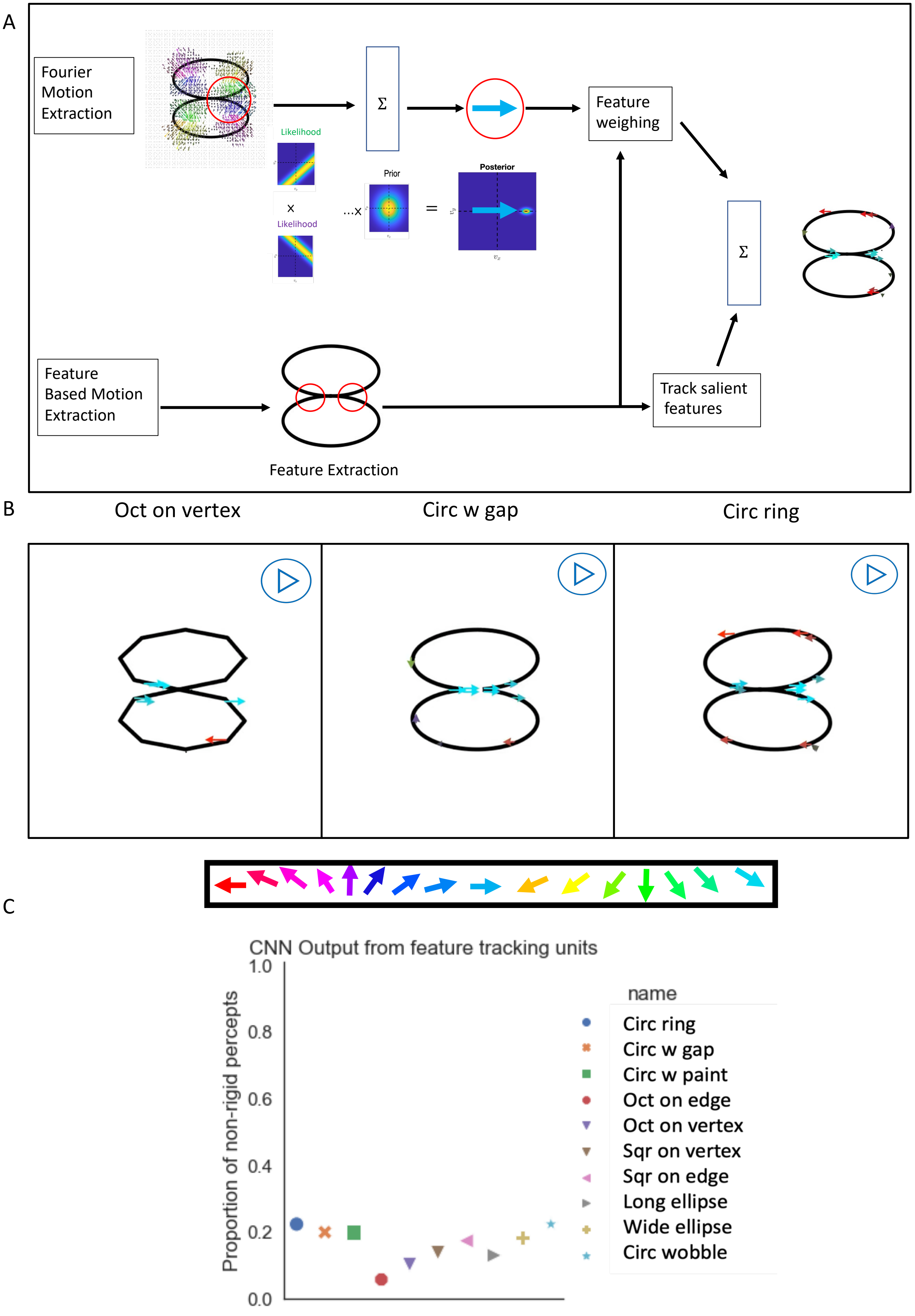

**Figure 7. Feature tracking mechanism.** (A): Two feature tracking streams simulating MT pattern-direction-selective (top) and feature-extraction based motion (bottom). Top: the inputs are the vectors attained from the motion energy units and each motion energy vector creates a likelihood function that is perpendicular to the vector. Likelihood functions are combined with the slowest motion prior by Bayes' rule. The output vector at each location is selected by the maximum posteriori. Bottom: salient features (corners) are extracted, and a velocity of the feature is computed by picking up the highest correlated location in the succeeding image. The outputs from two streams are combined. (B): The preferred velocity of the most responsive unit at each location is shown by a colored vector using the same key as Figure 6. Stimulus name is indicated on the bottom of each section. Most of the motion vectors point right or left, corresponding to the direction of rotation. (C): Average CNN output based on the feature tracking vector fields being the inputs for different stimulus shapes, shows higher probability of rigid percepts.

**Feature-tracking computations**

Motion energy signals support wobbling for a rigidly rotating ring-pair because the preferred direction of local motion detectors with limited receptive field sizes is normal to the contour instead of along the direction of object motion, known as the aperture problem (Stumpf, 1911; Wallach, 1935; Todorović, 1996; Wuerger et al., 1996; Bradley and Goyal, 2008). In many cases, the visual system resolves this ambiguity by integrating local velocities (Adelson and Movshon, 1982; Heeger, 1987; Recanzone et al., 1997; Simoncelli and Heeger, 1998; Weiss et al., 2002; Rust et al., 2006) or tracking specific features (Stoner and Albright, 1992; Wilson et al., 1992; Pack et al., 2003; Shiffrar and Pavel, 1991; Lorenceau and Shiffrar, 1992; Ben-Av and Shiffrar, 1995; Lorenceau & Shiffrar, 1999). In fact, humans can sometimes see unambiguous motion of shapes without a consistent motion energy direction by tracking salient features (Cavanagh and Anstis, 1991; Lu and Sperling, 2001), e.g. where the features that segment a square from the background such as gratings with different orientations and contrasts change over time while the square moves laterally. There is no motion-energy information that can support the movement of the square, yet we can reliably judge the direction of the movement. Consequently, to understand the contribution of salient features to percepts of rigidity, we built a feature tracking network.

Figure 7A shows a schematic diagram of the network. We used two ways to extract the direction of pattern motion, tracking of extracted features (Lu and Sperling, 2001; Sun et al., 2015) and motion energy combined into units resembling MT pattern-direction selective cells (Adelson and Movshon, 1982; Simoncelli et al., 1991; Weiss et al., 2002; Rust et al., 2006).

In the first module of the model (Bottom of Figure 7), features such as corners and sharp curvatures are extracted by the Harris Corner Detector (Harris and Stephens, 1988), but could also be extracted by end-stopped cells (Hubel and Wiesel, 1965; Dobbins et al., 1987; Pasupathy and Connor, 1999; Pack et al., 2003; Rodríguez-Sánchez and Tsotsos, 2012). The extracted features are tracked by taking the correlation between successive images and using the highest across-image correlated location to estimate the direction and speed of each feature's motion. The process is explained in equations in the **Appendix: Feature-tracking**.

The second module in feature-tracking combines the outputs from motion energy units into pattern-direction-selective (PDS) mechanisms (Top of Figure 7). The red circle indicates the receptive field of one pattern-selective unit subtending 0.9 degrees. The vectors within the receptive field are the motion-energy outputs that are inputs of the unit. The vectors emerging from the upper ring point down and to the right, indicated in green, and illustrating the aperture problem caused by a small receptive field size. The ambiguity created by the aperture problem is represented by the upper-panel likelihood function, with the probability distribution over the velocity components in horizontal and vertical directions ($V_x$ and $V_y$) represented by a heatmap with yellow indicating high probability and blue low probability velocities. The spread in the high-likelihood yellow region is similar to Wallach's illustration using two infinitely long lines (Wallach, 1935). The lower-panel likelihood represents the aperture problem from motion-energy vectors from the bottom ring. To model an observer who estimates pattern velocity with much less uncertainty, the local uncertain measurements are multiplied together by a narrow prior for the slowest motion, to obtain a velocity that maximizes the posterior distribution (MAP) (Simoncelli et al., 1991; Weiss et al., 2002). When the ring stimuli are run through arrays of pattern-direction selective cells, there is a response in the rotation direction at the joint and at some corners, but most of the responses are orthogonal to the contours as would be expected where locally there is a single contour (Zaharia et al., 2019) and would support wobbling classifications from the CNN. Feature tracking either requires attention, or is enhanced by it (Cavanagh, 1992; Lu and Sperling, 1995; Treue and Trujillo, 1999; Treue and Maunsell, 1999; Thompson and Parasuraman, 2012), so we use the corner detector to identify regions with salient features that could be tracked, and simulate the effect of attention by attenuating the gain of PDS units that fall outside windows of feature based attention. The attention-based attenuation suppresses pattern-selective responses to long contours and preferentially accentuates the motion of features such as corners or dots (Noest and Berg, 1993; Pack et al., 2004; Bradley and Goyal, 2008; Tsui et al., 2010).

At the last stage of Figure 7A, the vector fields from the pattern-selective units and feature tracking units are summed together. Figure 7B shows videos of samples of the feature tracking module's outputs for three examples of the ring stimuli, and the shape of the stimulus is shown at the bottom of each video. At the connection of the two rings, the preferred direction is mainly right or left, corresponding to the rotational direction. Depending on the phase of the rotation, the lateral velocities are also observed at sharp curvatures as the projected 2D contours deform significantly depending on the object pose angle (Koch et al., 2018; Maruya and Zaidi, 2020a and 2020b).

The combined vector fields were used as the inputs to the previously trained CNN for classification as rotating or wobbling. The results are shown in Figure 7C as probabilities of non-rigid classifications. The feature tracking vector fields generate classification proportions from 0.1 to 0.2 indicating rigidity, suggesting that feature tracking could contribute to percepts of rigidity in the rotating rings.

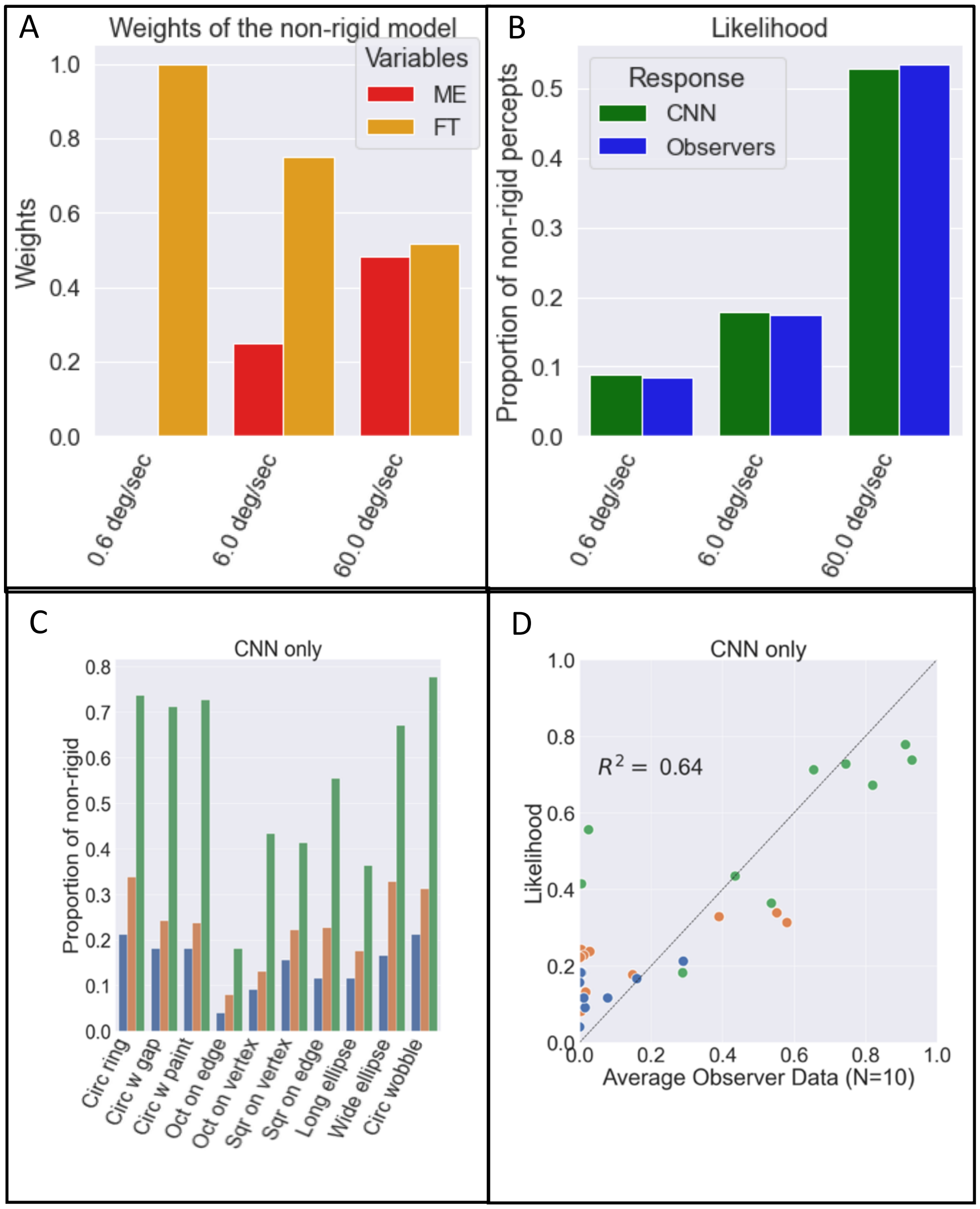

**Figure 8. Combining motion-energy and feature tracking outputs.** A: Estimated optimal weights of inputs from the motion energy mechanism (red) and the feature tracking mechanism (yellow) as a function of rotation speed overall shapes. B: CNN rigidity classifications as a function of rotation speed. The trained CNN output from the linear combination of two vector fields, the likelihood, is denoted by the green bar and the blue bar indicates the average of the 10 observers response. C: the proportion of non-rigid percept from the likelihood function of the CNN as a function of the speed of the stimulus for different shapes. Different colors show different speeds of stimulus (blue: 0.6 deg/sec, orange: 6.0 deg/sec, and green: 60.0 deg/sec). D: the likelihood of non-rigidity output plotted against average of observers' reports. At the fast speed, the model predicts similar probability of non-rigidity for shapes where the observers' percepts vary. Thus, the model doesn't capture the important properties of observer's percepts as a function of the shape of the object.

### Combining outputs of motion mechanisms for CNN classification

Psychophysical experiments that require detecting the motion direction of low contrast gratings superimposed on stationary grating pedestals have shown that feature tracking happens only at slow speeds and that motion energy requires a minimum speed (Lu and Sperling, 1995; Zaidi and DeBonet, 2000). We linearly combined the two vector fields attained from motion energy units and feature tracking units with a free weight parameter that was a function of speed, and the combined vector fields were fed to the trained CNN to simulate the observer's proportion of non-rigidity as a function of different shapes and different speeds (**Appendix: Combining motion mechanisms**). We tried weights between 0.0 to 1.0 every 0.01 increment at each speed to minimize the mean square error (MSE) from the observers' average. The optimal weights are shown in Figure 8A. The weight for the FT decreases and the weight for the ME increases as a function of rotation speed, consistent with earlier results on the two motion mechanisms. In Figure 8B, the green bars show that the average proportion of non-rigid classifications generated by the CNN output across speeds is very similar to the average percepts (blue bars). However, the average across all shapes hides the fact that the proportion of non-rigid classifications from the CNN only explains a moderate amount of variance in the proportions of non-rigid percepts if examined for separate shapes in Figures 8C & D ($R^2$= 0.64). The scattergram shows that the CNN classification is systematically different from the perceptual results at the fastest speed where the prediction is flat across different shapes while the observers' responses vary with shape. Next, we examine possible factors that could modify percepts as a function of shape.

### Priors and illusions

The video in Figure 1A shows that wobbling is not the only non-rigid percept when the circular rings are rotated, as the top ring also seems to be rolling around its center. Unlike the motion-energy support for wobbling, and the feature-tracking support for rotation, there are no motion-energy or feature-tracking signals that would support the rolling percept (Videos Figures 5D, E, and 7B), which would require local motion vectors tangential to the contours of the rings. To illuminate the factors that could evoke or stop the rolling illusion, we show a simpler 2D rolling illusion. In the video in Figure 9A, a circular untextured 2D ring translated horizontally on top of a straight line is perceived predominantly as rolling like a tire on a road. The perception of rolling would be supported by motion signals tangential to the contour (Figure 9B), but the local velocities extracted from the stimulus by an array of motion-energy units are predominantly orthogonal to the contour as expected from the aperture effect (Video in Figure 9C), while feature tracking extracts motion predominantly in the direction of the translation (Video in Figure 9D), both of which should counter the illusion of clockwise rolling. Hence the rolling illusion goes against the sensory information in the video. This illusion could demonstrate the power of prior probabilities of motion types. To identify factors that enhance or attenuate the illusion, we performed experiments on the two-ring configurations.

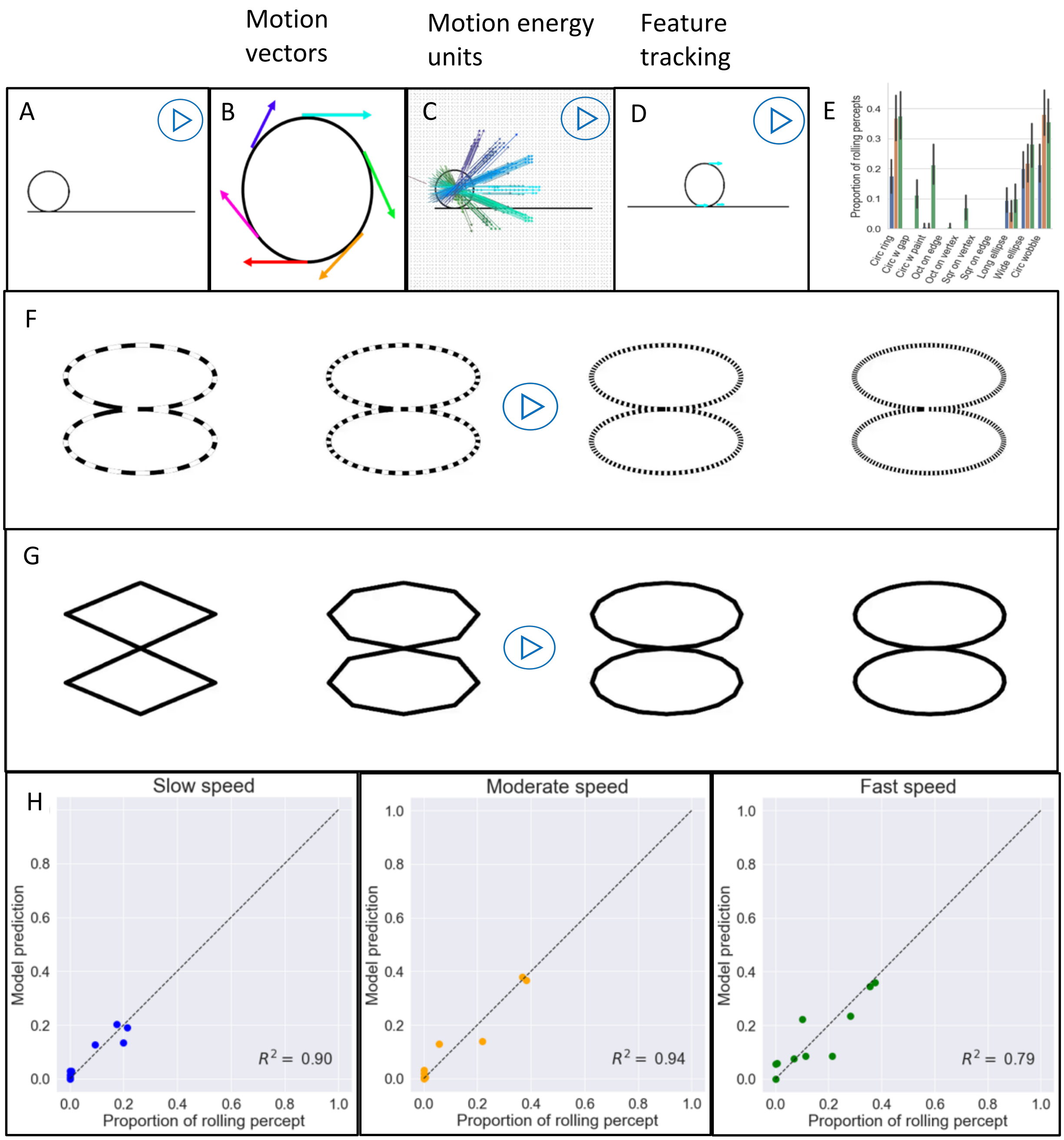

**Figure 9. Rolling illusion:** (A) shows a 2D circle on a line, translating from left to right. Our percept of the translating circle, however, is rolling clockwise. To perceive the rolling based on the sensory information, local motion units that direct tangential to the contour (B) are required. (C) and (D) show local motion selective units from motion energy (left) and feature tracking (right). In both cases, the vectors are inconsistent with the required vectors. (E): Average proportion of rolling percepts (8 observers). The color of the bar shows the different speed of stimulus (blue: 0.6 deg/sec, orange: 6.0 deg/sec, and green: 60.0 deg/sec). The shape of the stimulus is indicated on the x-axis. The proportion of rolling percepts increased with speed and decreased when features were added to the rings. (F): Rolling illusion and rotational symmetry. The non-rigidity (rolling) percepts increases with the order of rotational symmetry from left to right. (G): The relationship between rolling illusion and the strength of feature. As the number of corners increase from left to right, it gets harder to extract the corners and accordingly, the percept of rolling increases. H: Model prediction with rotational symmetry and average strength of features versus average proportion of rolling percepts for slow (left), moderate (middle), and fast (right) speeds ($R^2 = 0.90, 0.94, and\ 0.79$).

## Quantifying the rolling illusion

We quantified the perception of rolling in the original ring illusion by using the same ring pairs as in Figure 3 (0.6, 6.0, 60.0 deg/sec; 3 deg), but now on each trial the observers were asked to respond Yes or No to the question whether the rings were rolling individually around their own centers. The results are plotted in Figure 9E (for 20 repetitions/condition and 8 observers). The frequency of trials seen as rolling increased with speed and decreased when gaps, paint, or corners were added to the rings, with corners leading to the greatest decrease. We think this illusion demonstrates the power of prior expectations for rolling for different shapes. The prior probability for rolling could reflect the rotational symmetry of the shape, as circular rings with higher order rotational symmetries are more likely to be seen as rolling and wobbling in Figure 9F. In addition, the more features such as corners that are seen as not rolling, the more they may attenuate the illusion as shown in Figure 9G. It's possible that aliasing increases with the degree of symmetry at high speeds and features may be less effective at high speeds if they get blurred in the visual system, explaining greater rolling and wobbling at higher speeds. The degree of rotational symmetry of a circle is infinite, so we reduced it to the number of discrete pixels in the circumference and regressed the proportion of rolling percepts against the log of the order of rotational symmetry of each shape and the mean strength of features, $\bar{h}$ (the average value of $h$ defined in equation (A18)). The two factors together predicted rolling frequency with $R^2$ = 0.90, 0.94 & 0.79 for slow, medium, and fast speeds (Figure 9H).

## Adding prior assumptions to motion mechanism-based CNN classification for rigid and non-rigid perception of the rotating ring-pairs

The first model showed that to completely explain the variation of the illusion where rigid rings are perceived as non-rigidly connected as a function of speed and shape, other factors have to be considered besides the outputs of bottom-up motion mechanisms. The degree of rotational symmetry may supply not only a prior for rolling but also for wobbling as *a priori* a circular ring is more likely to wobble than a square ring. We thus set up a prior dependent on the number of rotational symmetries and the average strength of the detected features. The posterior probability of a motion class is thus conditional on the motion vector fields, the speed which determines the relative weights of the motion energy and feature tracking motion fields, and the shape-based priors. These factors are combined with weights and the posterior is computed by using Bayes' rule (**Appendix: Final model**). When the weights were estimated by using gradient descent to minimize the mean square error, the three factors together predicted proportions of non-rigid percepts with $R^2$ = 0.95 for all the speeds combined (Figure 10).

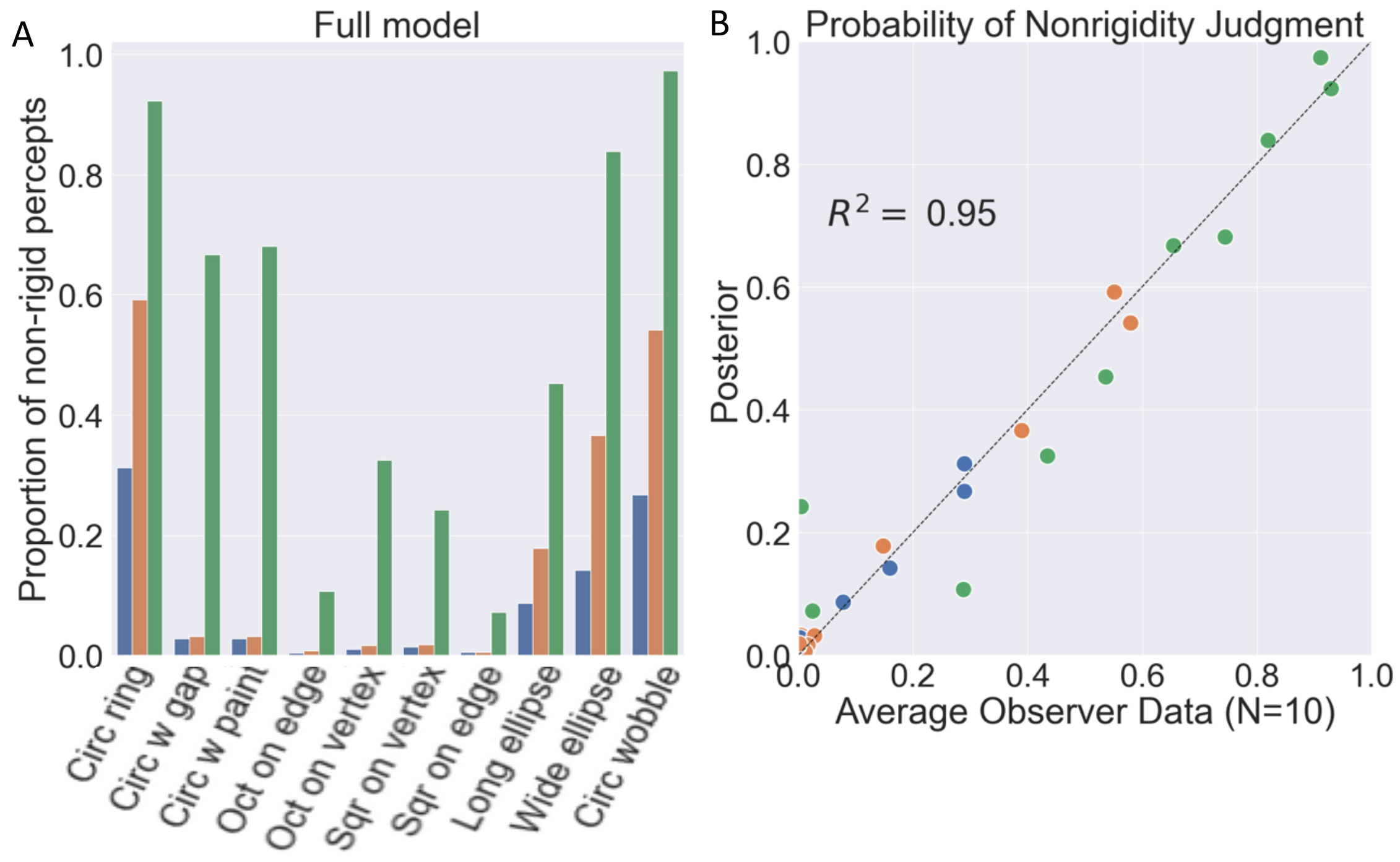

Figure 10. **Final model:** (A): the proportion of non-rigid percept classifications as a function of the speed of the stimulus and different shape from the final model combining shape-based rolling and wobbling priors with the CNN likelihoods. Different colors show different speeds of the stimulus (blue: 0.6 deg/sec, orange: 6.0 deg/sec, and green: 60.0 deg/sec). (B): the posterior output plotted against the observer's percepts. The model explains the observer's percepts ($R^2$: 0.95).

**Discussion**

We began this article by stating that our aim was to bridge the gap between understanding neuronal stimulus preferences and understanding how cooperation and competition between different classes of neuronal responses generates visual perception, and to identify the factors that govern transitions between stable states of perception. We now discuss how we have attempted this and how far we have succeeded.

We used a variant of a physical illusion that has been widely used but never explained. A pair of circular rings rigidly connected appear non-rigidly connected when rotated. By varying rotation speed, we discovered that even for physical objects, percepts of rigidity dominate at slow speeds and percepts of non-rigidity at high speeds. We then varied shapes using computer graphics and discovered that the presence of vertices or gaps or painted segments promoted percepts of rigid rotation at moderate speeds where the circular rings looked nonrigid, but at high speeds all shapes appeared nonrigid.

By analyzing the stream of images for motion signals, we found that the non-rigid wobble can be explained by the velocity field evoked by the rigidly rotating object in an array of motion energy units, because of the limited aperture of the units. This explanation of course depends on the decoding of the motion field. In the absence of knowledge of biological decoders for object motion, it is possible to infer degree of rigidity from the velocity field itself. Analytic tests of the geometrical rigidity of velocity fields can be based on the decomposition of the velocity gradient matrix, or on an analysis of the temporal derivative of the curvature of moving plane curves, but limitations of both approaches have been noted (Todorović, 1993). We did not attempt to recreate the complete percept or to analyze elastic versus articulated motion (Jasinschi, & Yuille, 1989 ; Aggarwal et al., 1998, Jain & Zaidi, 2011), but restricted the analysis to distinguish wobbling from rigid rotation. For this purpose, we trained a CNN which could make this distinction for many velocity fields. The CNN indicated an almost 100% confidence in wobble from the velocity fields of all shapes and at all speeds, pointing to the need for other mechanisms that take speed and shape into account to explain the results.

The obvious second mechanism to explore was feature tracking, whose output can depend on the salience of features. For this purpose, we used the Harris corner detector that is widely used in computer vision, but also pattern motion selective units. The velocity field from this mechanism was judged by the trained CNN to be compatible with rigid rotation, with little variation based on shape. The output of the CNN can be considered to classify the information present in a velocity field without committing to a particular decoding process. By making the empirically supported assumption that motion energy needs a minimum speed and that feature tracking only functions at slow speed, the CNN output from combined two vector fields could be combined to explain the empirical results with an $R^2$=0.64, mainly by accounting for the speed effects. An inspection of empirical versus predicted results showed the need for mechanisms or factors that were more dependent on shape.

In both the physical and graphical stimuli, observers see an illusion of the rings rolling or spinning around their own center. This illusion is remarkable because there are no sensory signals that support it, shown starkly by translating a circular ring along a line. The illusion is however suppressed by vertices, gaps and painted segments, suggesting that a powerful prior for rolling may depend on rotational symmetry or jaggedness of the shape. The addition of this shape-based prior to the model, leads to an $R^2$=0.95, which suggests that we have almost completely accounted for the most important factors.

To summarize, we show how visual percepts of rigidity or non-rigidity can be based on the information provided by different classes of neuronal mechanisms, combined with shape-based priors. We further show that the transition from perception of rigidity to non-rigidity depends on the speed requirements of different neuronal mechanisms.

## APPENDIX

### Stimulus generation and projection

We generated 3D rotating, wobbling, and rolling stimuli by applying the equations for rotation along each of $X$ , $Y$, $and$ $Z$ axes in a 3-D space to all points on the rendered objects. The rotational matrix around each axis $R_{axis}(\theta)$ is:

$$R_X(\theta) = \begin{pmatrix} 1 & 0 & 0 \\ 0 & \cos\theta & -\sin\theta \\ 0 & \sin\theta & \cos\theta \end{pmatrix} \quad \cdots (A1)$$

$$R_Y(\theta) = \begin{pmatrix} \cos\theta & 0 & \sin\theta \\ 0 & 1 & 0 \\ -\sin\theta & 0 & \cos\theta \end{pmatrix} \quad \cdots (A2)$$

$$R_Z(\theta) = \begin{pmatrix} \cos\theta & -\sin\theta & 0 \\ \sin\theta & \cos\theta & 0 \\ 0 & 0 & 1 \end{pmatrix} \quad \cdots (A3)$$

If $\vec{O}$ is the initial position of a 3-D point on the object lying on $XY$ plane, the location of the point ($\overrightarrow{O_{rot}}$) on a rotating object inclined at an angle of $\phi$ from the ground plane and angular velocity of $\omega$ is expressed by equation A4:

$$\overrightarrow{O_{rot}} = R_Z(\omega t) R_Y(\phi)\, \vec{O} \quad \cdots (A4)$$

The wobbling object is described by A5:

$$\overrightarrow{O_{wbl}} = R_Z(\omega t) R_Y(\phi)\, R_Z(-\omega t) \vec{O} \quad \cdots (A5)$$

The rolling object by A6:

$$\overrightarrow{O_{rll}} = R_Z(\omega t) R_Y(\phi)\, R_Z(\omega t) \vec{O} \quad \cdots (A6)$$

The projected image $\vec{I}$ of the $x, y, and\ z$ components of $\overrightarrow{O_{mov}}$ to the screen at the time $t$ is calculated by A7:

$$\vec{I} = \begin{pmatrix} \dfrac{x f_c}{d_c - y} \\ \dfrac{z f_c}{d_c - y} \\ t \end{pmatrix} \quad \cdots (A7)$$

Where $f_c$ is the focal length of the camera, and $d_c$ is the distance from the camera to the object.

The difference in the equations between rotation and wobbling or rolling are the rotations around the center of the object: $R_Z(-\omega t)\vec{O}$ or $R_Z(\omega t)\vec{O}$ . These rotations are not discernible for circular rings, so rotating, wobbling, and rolling circular rings generate the same images on the screen.

**Motion-energy**

To understand the response of the motion-energy mechanism to the rotating rings, we generated arrays of motion-energy filters that were convolved with the video stimulus. Each filter was based on a pair of odd and even symmetric Gabors in quadrature pair. We first computed the i-th pair of Gabor filters at each position and time:

$$G_{i,odd}(x,y,t) = \exp\left(-\frac{(x-c_{x,i})^2+(y-c_{y,i})^2}{2{\sigma_{s,i}}^2} - \frac{\left(t-c_{t,i}\right)^2}{2{\sigma_{t,i}}^2}\right) \times \sin\left(\left(x-c_{x,i}\right)f_{x,i} + \left(y-c_{y,i}\right)f_{y,i} + (t-c_{t,i})f_{t,i}\right) \quad \cdots (A8)$$

$$G_{i,even}(x,y,t) = \exp\left(-\frac{(x-c_{x,i})^2+(y-c_{y,i})^2}{2{w_{s,i}}^2} - \frac{\left(t-c_{t,i}\right)^2}{2{w_{t,i}}^2}\right) \times \cos\left(\left(x-c_{x,i}\right)f_{x,i} + \left(y-c_{y,i}\right)f_{y,i} + (t-c_{t,i})f_{t,i}\right) \quad \cdots (A9)$$

Where $c_{x,i}$, $c_{y,i}$ are the center of the filter in space and $c_{t,i}$ is the center in time; $\sigma_{s,i}$ and $\sigma_{t,i}$ are the spatial and temporal standard deviations of the Gaussian envelopes, and $f_{x,i}$, $f_{y,i}$, and $f_{t,i}$ are the spatial and temporal frequency of the sine component of the Gabor, referred to as the preferred spatial and temporal frequencies. Each filter $G_i$ was convolved with the video $I(x,y,t)$, then the responses of the quadrature pair were squared and added to give a phase independent response $ME_i(x,y,t)$. At each location $(x,y,t)$, the preferred spatial and temporal frequencies, $\widehat{f_x}, \widehat{f_y}, and\ \widehat{f_t}$ of the filter that gave the maximum response $ME_i(x,y,t)$ was picked:

$$\widehat{f_x}, \widehat{f_y}, \widehat{f_t} = \underset{f_{i,x}, f_{i,y}, f_{i,t}}{\operatorname{argmax}} \left(ME_i(x,y,t)\right) \quad \cdots (A10)$$

Then, the speed, $S(x,y,t)$, and the direction of the velocity, $\theta(x,y,t)$, were calculated as:

$$S(x, y, t) = \frac{\widehat{f_t}}{\sqrt{\widehat{f_x}^2 + \widehat{f_y}^2}} \qquad \cdots (A11)$$

$$\theta(x, y, t) = \arctan\left(\frac{\widehat{f_y}}{\widehat{f_x}}\right) \qquad \cdots (A12)$$

Thus, the vector field, $Q_{ME}$, attained from motion energy units in the horizontal and vertical components of the velocity, $Q_{ME}$, will be:

$$Q_{ME}(x, y, t) = \begin{pmatrix} S(x, y, t) * cos\theta(x, y, t) \\ S(x, y, t) * sin\theta(x, y, t) \end{pmatrix} \qquad \cdots (A13)$$

**Convolutional Neural Network (CNN)**

The CNN has two convolutional layers and one fully-connected layer with the softmax activation function. For the first convolutional layer, each vector field at time $t$ is cross-correlated with 32 filters with the size of $3 \times 3$ and a non-linear rectification is applied:

$$a_i^1 = Max_i\left(ReLU\left(\sum_{c,m,n} w_i^{1,c}[x+m, y+n]Q_S[x, y, t|V_S] + b_i^{1,c}\right)\right) \qquad \cdots (A14)$$

Where *c* is the horizontal and vertical components of the velocity, $w_i^c$ is the $i$th weight of 32 filters, $b_i^c$ is the $i$th bias, $ReLU(x) = \max(x, 0)$ is the rectified linear activation function, and $Max_i$ is the max-pooling with the size of $2 \times 2$. Then, in the second layer, each output of $a_i^1$ is cross-correlated with 10 $3 \times 3$ filters followed by the $ReLU$ function:

$$a_{i,j}^2 = ReLU\left(\sum_{m,n} w_j^2[x+m, y+n]a_i^1[x, y] + b_j^2\right) \qquad \cdots (A15)$$

Finally, $a_{i,j}^2$ is flattened to be a vector, $\overrightarrow{a^2}$, and the output of the CNN is computed by the fully-connected layer with the softmax activation function:

$$f_{CNN}^t = softmax(W^3\overrightarrow{a^2} + b^3) \qquad \cdots (A16)$$

Where the softmax activation function is calculated by:

$$softmax(Z)_i = \frac{e^{z_i}}{\sum e^{z_j}} \qquad \cdots (A17)$$

**Feature-tracking**

Our Feature-tracking mechanism simulation has two modules. In the first, module, at corners and sharp curvatures, the image intensity changes along different directions, and the Harris Corner Detector exploits this property to extract salient features. First, we computed the change in intensity value of a part of image by sifting a small image patch in all directions and taking a difference between the patch and the shifted one. Suppose that $I(x, y)$ is the image intensity at $(x, y)$ position and consider a small image patch (receptive field) with a Gaussian window, $(x, y) \in W$ ($W$ is a Gaussian window, of which the size is $5 \times 5$ (pixels)). If the window is shifted by $(\Delta x, \Delta y)$, the sum of the squared difference (SSD) between two image patches will be:

$$E(\Delta x, \Delta y) = \sum_{x,y} W(x, y)\,(I(x, y) - I(x + \Delta x, y + \Delta y))^2$$

From the first order Taylor expansion,

$$I(x + \Delta x, y + \Delta y) \approx I(x, y) + I_x(\mathrm{x}, \mathrm{y})\Delta x + I_y(\mathrm{x}, \mathrm{y})\Delta y \qquad \cdots (A18)$$

The SSD will be approximated by:

$$\begin{aligned} E(\Delta x, \Delta y) &\approx \sum_{x,y} W(x, y)\,(I(x, y) - (I(x, y) + I_x(\mathrm{x}, \mathrm{y})\Delta x + I_y(\mathrm{x}, \mathrm{y})\Delta y))^2 \\ &= \sum_{x,y} W(x, y)\,(I_x(\mathrm{x}, \mathrm{y})\Delta x + I_y(\mathrm{x}, \mathrm{y})\Delta y)^2 \end{aligned} \qquad \cdots (A19)$$

Which can be written in matrix form:

$$\begin{aligned} &= (\Delta x \quad \Delta y) \sum_{x,y} W(x, y) \begin{pmatrix} I_x^2 & I_x I_y \\ I_x I_y & I_y^2 \end{pmatrix} \begin{pmatrix} \Delta x \\ \Delta y \end{pmatrix} \\ &= (\Delta x \quad \Delta y) M \begin{pmatrix} \Delta x \\ \Delta y \end{pmatrix} \\ &= (\Delta x \quad \Delta y) U \Lambda U^T \begin{pmatrix} \Delta x \\ \Delta y \end{pmatrix} \end{aligned} \qquad \cdots (A20)$$

Where $U$ is an orthonormal matrix containing the two eigenvectors of $M$ and $\Lambda$ is a diagonal matrix, $\Lambda = \begin{pmatrix} \lambda_1 & 0 \\ 0 & \lambda_2 \end{pmatrix}$, with the two eigenvalues. Eigenvalues quantify change in the image intensity values along the eigenvectors and they differ based on image properties. Figure A1 shows those properties. At a uniform region on the left, there is no change in the image intensity at the receptive field in accordance with the displacement of the receptive field ($\lambda_1$ and $\lambda_2$ are close to zero). When the receptive field is close to the edge, the image intensity changes in a direction perpendicular to the contour, but not along the contour ($\lambda_1$ has a large value, but $\lambda_2$ is small). However, at the corner, the image intensity differs in all directions (both $\lambda_1$ and $\lambda_2$ have a large value). We quantified features (corners) by the following equation:

$$h = \lambda_1\lambda_2 - K(\lambda_1 + \lambda_2)^2 = \det(\Lambda) - K\, tr(\Lambda)^2 \qquad \cdots (A21)$$

$tr$ is the trace of the matrix, and the threshold $K$ is set at 0.05 as a default by MATLAB. We extracted local features if $h > 0.05$. Suppose that the extracted local feature that satisfies $h > 0.05$ is $L_F(cx_t, cy_t)$ where $cx_t$ $and$ $cy_t$ are the center of a 5x5 pixel extracted feature at time $t$, then this local feature is cross correlated with the succeeding frame to extract the next location of the feature:

$$cx_{(t+1)}, cy_{(t+1)} = \underset{x,y}{\operatorname{argmax}}\Big(\sum_u \sum_v I(u, v, t+1) L_F(x+u, y+v)\Big) \qquad \cdots (A22)$$

Then, the vector fields $Q_H$ are extracted by:

$$Q_H(cx_t, cy_t, t) = \begin{pmatrix} cx_{(t+1)} - cx_t \\ cy_{(t+1)} - cy_t \end{pmatrix} \qquad \cdots (A23)$$

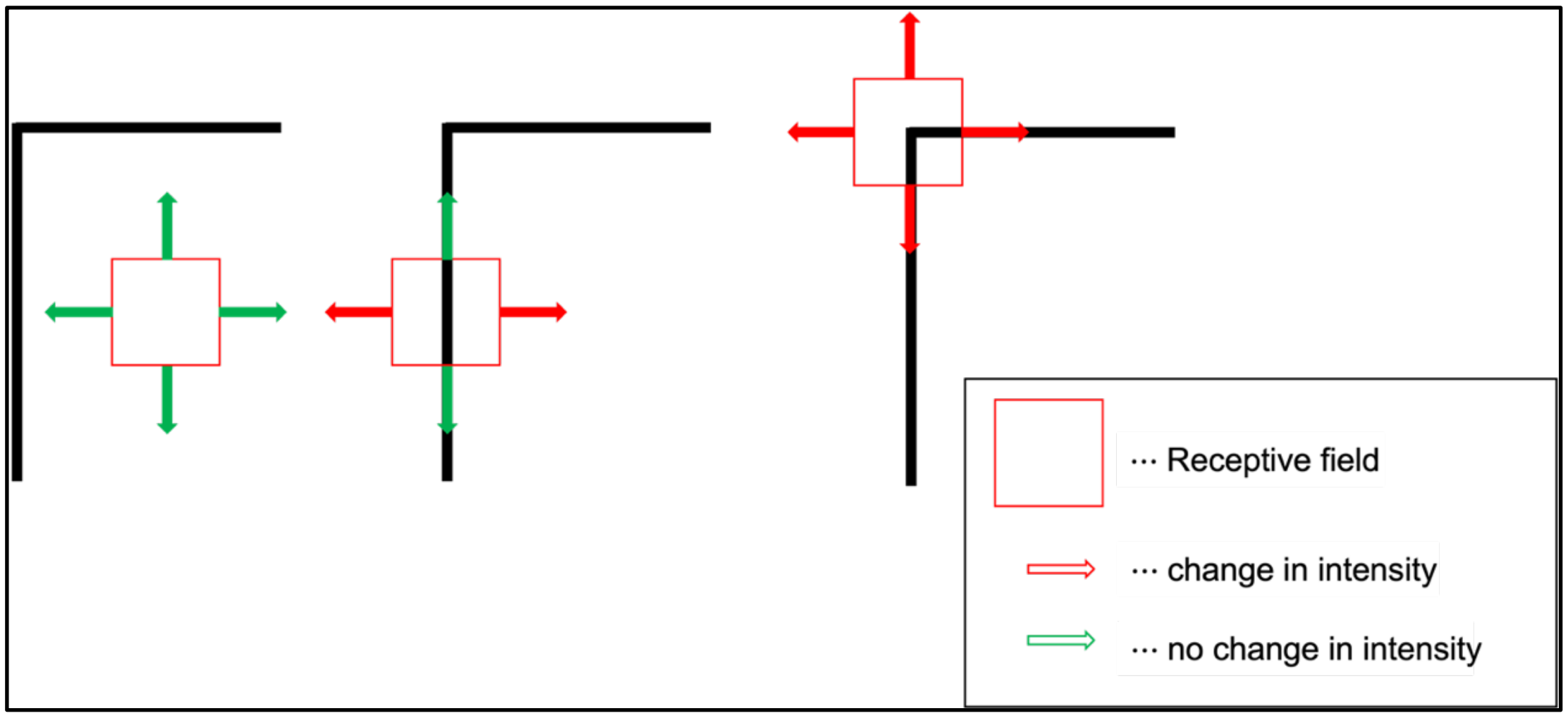

**Figure A1. Feature selection.** The change in the image intensity across different regions. The red square shows a receptive field at a flat region (left), at an edge (middle), and at a corner (right). For the flat region, a small change in the receptive field location doesn't change the image intensity (shown in green arrows). At the edge, moving along the edge direction doesn't change the overall image intensity, but except for that direction, the image intensity shifts especially along the direction perpendicular to the edge. At the corner, the overall image intensity changes in every direction.

The second module in feature-tracking combines the outputs from motion energy units into pattern-direction-selective (PDS) mechanisms (Top of Figure 7A). The input of this unit is the motion energy vector flow, given by Equations A11 and A12. Considering only the locations $(cx_t, cy_t)$, the center of locations extracted from the Harris Corner Detector, for each vector, the ambiguity resides along a perpendicular line from the vector, and we assumed the observer's measurement is ambiguous. Thus, the $i\ th$ measurement distribution given the local velocity, $\vec{v}$, is represented by a Gaussian distribution:

$$p(S_i, \Theta_i | \vec{v}(cx_t, cy_t, t)) \propto \exp\left(-\frac{1}{2\sigma^2}(sin\Theta_i v_x + cos\Theta_i v_y - S_i)^2\right) \quad \cdots (A24)$$

Then, we combine the observer's measurement with the slowest motion prior, represented as following:

$$p(\vec{v}) \propto \exp\left(-\frac{v_x^2 + v_y^2}{2{\sigma_p}^2}\right) \quad \cdots (A25)$$

Where $\sigma_p$=2 (arbitrarily chosen). The posterior probability over a velocity would be computed by using Bayes' rule:

$$p(\vec{v}|S_1, \Theta_1, \dots, S_n, \Theta_n) \propto p(\vec{v})p(S_1, \Theta_1, \dots, S_n, \Theta_n|\vec{v})$$

By assuming conditional independence, $S_1, \Theta_1 \perp S_2, \Theta_2 \ \dots \perp S_n, \Theta_n | \vec{v}$, the equation becomes:

$$p(\vec{v}|S_1, \Theta_1, \dots, S_n, \Theta_n) \propto p(\vec{v}) \prod_i^n p(S_i, \Theta_i|\vec{v}) \quad \cdots (A26)$$

Then, the local velocity was estimated by taking the maximum posteriori, so the velocity field from PDS becomes:

$$Q_{PDS}(cx_t, cy_t, t) = \underset{\vec{v}}{\operatorname{argmax}}(p(\vec{v}) \prod_i^n p(S_i, \Theta_i|\vec{v})) \quad \cdots (A27)$$

The vector fields $Q_{FT}$ from these two feature tracking units are combined by a vector sum.

$$Q_{FT} = Q_{ME} + Q_{PDS} \quad \cdots (A28)$$

**Combining motion mechanisms**

We linearly combined the motion energy and feature tracking vector fields with different combinations of weights that summed to 1.0 and the combined vector fields were fed to the trained CNN to classify the observer's proportion of non-rigidity as a function of different speeds. Suppose $Q_S$ is the combined field and $V_S$ is the speed of the stimulus. $Q_S$ is computed by the weighted sum of two velocity field such that:

$$Q_S(x, y, t|V_S) = w(V_S)Q_{ME} + \left(1 - w(V_S)\right)Q_{FT} \quad \cdots (A29)$$

$w(V_S)$ is a weight function that depends on the speed of stimulus. The higher $w(V_S)$ is, the more the vector field gets closer to $Q_{ME}$. Conversely, if $w(V_S)$ is lower, the vector field resembles more to the rigid rotation. The likelihood of $C_M$, classification of motion as wobbling or rotation, is estimated by the trained CNN as a function of $Q_S$ and $V_S$:

$$L(C_M) = p(Q_S|C_M, V_S, S_h)$$

$$= \bar{f}_{CNN}(Q_S) \qquad \cdots (A30)$$

Where $\bar{f}_{CNN}$ is computed by the average of $f_{CNN}^{t}$ (Eq. A16) across all time.

**Final model**

The first model showed that to completely explain the variation of the illusion where rigid rings are perceived as non-rigidly connected as a function of speed and shape, other factors have to be considered besides the outputs of the two bottom-up motion mechanisms. In this section we add prior assumptions to motion mechanism-based CNN classifications for rigid and non-rigid perception of the rotating ring-pairs. The degree of rotational symmetry may supply not only a prior for rolling but also for wobbling as *a priori* a circular ring is more likely to wobble than a square ring. Suppose that $S_h = \binom{n_s}{\bar{h}}$ where the number of rotational symmetries is $n_s$, and the average strength of the detected corner is $\bar{h}$. The posterior probability of a motion class, $C_M$, given the vector fields, rotation speed, and object shape is computed by using Bayes' rule:

$$p(C_M|Q_S, V_S, S_h) = \frac{p(C_M)p(Q_S, V_S, S_h|C_M)}{p(Q_S, V_S, S_h)}$$

By factorizing the conditional probability:

$$p(C_M|F_S, V_S, S_h) = \frac{p(C_M)p(V_S|C_M)p(S_h|C_M, V_S)p(Q_S|C_M, V_S, S_h)}{p(Q_S, V_S, S_h)}$$

$$p(C_M|F_S, V_S, S_h) = \frac{p(C_M)p(V_S|C_M)p(S_h|C_M, V_S)\bar{f}_{CNN}(Q_S)}{p(Q_S, V_S, S_h)}$$

$$= \frac{p(C_M)p(V_S, C_M)p(S_h, C_M, V_S)\bar{f}_{CNN}(Q_S)}{p(C_M)p(C_M, V_S)p(Q_S, V_S, S_h)}$$

$$= \frac{p(C_M|S_h, V_S)p(S_h, V_S)\bar{f}_{CNN}(Q_S)}{p(Q_S|S_h, V_S)p(S_h, V_S)}$$

$$= \frac{p(C_M|S_h\,,V_S)\bar{f}_{CNN}(Q_S)}{p(Q_S|S_h\,,V_S)} \qquad \cdots (A31)$$

Since the strength of features and the rotational symmetry seemed to be related to the percept of non-rigidity/rolling as shown in the rolling illusion experiment despite that there is no vector field that supports it, the conditional prior $p(C_M|S_h\,,V_S)$ is estimated by the following equation:

$$p(C_M|S_h\,,V_S) = \varsigma\left(\vec{W}(V_S)^T\,S_h + b(V_S)\right) \qquad \cdots (A32)$$

Where $b(V_S)$ and $\vec{W}(V_S)$ are a $2 \times 1$ weight vector and bias, both of which are dependent on the speed of the stimulus. $\varsigma$ is a sigmoid function:

$$\varsigma(x) = \frac{1}{1 + e^{-x}} \qquad \cdots (A33)$$

Thus, the posterior becomes:

$$p(C_M|Q_S, V_S, S_h\,) = \alpha\varsigma\left(\vec{W}(V_S)^T\,S_h + b(V_S)\right) \times \bar{f}_{CNN}(Q_S) \qquad \cdots (A34)$$

Where $\alpha$ is proportional to $p(Q_S|S_h\,,V_S)$ thus depending on the speed of the rotation. $\vec{W}$, $b$, and $\alpha$ are estimated by using gradient descent to minimize the MSE.